# Observation of polaronic state assisted sub-bandgap saturable absorption

Li Zhou[1], Yiduo Wang[1,2], Jianlong Kang[1], Xin Li[1], Quan Long[1], Xianming Zhong[1], Zhihui Chen[1], Chuanjia Tong[1], Keqiang Chen[3], Zi-Lan Deng[4], Zhengwei Zhang[1], Chuan-Cun Shu[1], Yongbo Yuan[1], Xiang Ni[1], Si Xiao[1], Xiangping Li[4], Yingwei Wang[1]*, Jun He[1]*.

[1]Hunan Key Laboratory of Nanophotonics and Devices, School of Physics, Central South University, Changsha 410083, China

[2]Beijing National Laboratory for Condensed Matter Physics, Institute of Physics, Chinese Academy of Sciences, Beijing 100190, China

[3]Faculty of Materials Science and Chemistry, China University of Geosciences, Wuhan 430074, China

[4]Guangdong Provincial Key Laboratory of Optical Fiber Sensing and Communications, Institute of Photonics Technology, College of Physics & Optoelectronic Engineering, Jinan University, Guangzhou 510632, China.

*e-mail: wyw1988@csu.edu.cn; junhe@csu.edu.cn

**Abstract**

Polaronic effects involving stabilization of localized charge character by structural deformations and polarizations have attracted considerable investigations in soft-lattice lead-halide perovskites. However, the concept of polaron assisted nonlinear photonics remains largely unexplored, which has a wide range of applications from optoelectronics to telecommunications and quantum technologies. Here, we report the first observation of the polaronic state assisted saturable absorption through sub-bandgap excitation with a redshift exceeding 60 meV. By combining photoluminescence, transient absorption measurements and density functional theory calculations, we explicate that the anomalous nonlinear saturable absorption is caused by the transient picosecond timescale polaronic state formed by strong carrier/exciton-phonon coupling effect. The bandgap fluctuation can be further tuned through exciton-phonon coupling of perovskites with different Young's modulus. This suggests that we can design targeted soft lattice lead-halide perovskite with a specific structure to effectively manipulate exciton-phonon coupling and exciton-polaron formation. These findings profoundly expand our understanding of exciton-polaronic nonlinear optics physics and provide an ideal platform for developing actively tunable nonlinear photonics applications.

**Introduction**

Nonlinear optics, describing how light-matter interaction depends upon the intensity of light filed in a nonlinear manner [1], has offered an unparalleled approach for quantum material and information[2, 3], all optical elements[4] or integrated photonic devices[5-7]. Nonlinear saturable absorption (SA) response as a typical nonparametric third-order nonlinear optical process (see details in Supplementary Note 1) has led to various promising applications in ultrashort pulse generation[8], and optical diodes[9]. In addition, the compelling demands for superior optical nonlinearity and active controlling nonlinear optical response of two-dimensional materials have further motivated intense research on exploring nonlinear optical modulation strategy including electric field regulation[10], excitation resonance[11] and heterojunction construction[12] etc. Substantial efforts have been devoted toward investigating polarons in two-dimensional material systems[13], with the aim of harnessing their potential as a solid-state medium for achieving enhanced nonlinear optical response[14, 15].

Polarons, as quasiparticles arising from strong interaction between electrons and lattice vibrations of material, have enabled versatile control of materials' functionalities. Nontrivial polarization in exciton-polarons was induced by strong coupling of long-lifetime photoexcited carrier/exciton with local lattice distortions in material[16, 17]. Compared with traditional semiconductors, the soft and polar lattice of lead halide perovskites (LHPs) usually exhibit large anharmonicity and dynamic disorder, leading to strong electron-phonon coupling[18, 19]. The polarons model proposed in perovskites has impelled widespread research interests ranging from polaron formation photophysical mechanism[16, 20], polarons dispersion relation detection[21, 22], to polarons contributed optoelectronic applications (solar cells[23, 24], laser[25], and photodetectors[26]). Efforts to understand exciton-polarons physics in soft lattice LHPs revealed that exciton polaronic feature enable enhanced sub-bandgap linear optical absorption and emission[27, 28]. Very recently, polaronic states leading to giant anti-Stokes shift up to 220 meV have been achieved in two-dimensional perovskite systems[29], demonstrating the importance of exciton–phonon coupling in tailoring interplay between soft-lattice dynamics and optoelectronic properties in perovskite material systems. However, a direct signature of the correlation between polaronic states and nonlinear optical responses in LHPs has yet to be reported. Achieving this observation is crucial for revealing how exciton-phonon coupling influences nonlinear nanophotonic responses, which holds significant implications for optoelectronics and quantum technologies.

In this article, we demonstrate polaronic state assisted nonlinear saturable absorption (PSA_SA) response in perovskites by observing nonlinear optical bleaching under sub-bandgap excitation at room temperature. Our systematic multi-wavelength Z-scan measurements show that the PSA_SA can be observed in $CsPbBr_3$ under sub-bandgap excitation with redshifts up to 60 meV. Our robust photoluminescence (PL) and transient absorption (TA) spectroscopy studies unequivocally confirm that this anomalous SA is caused by the polaronic state formed by strong carrier/exciton-phonon coupling effect, that manifests by phonon-assisted up-conversion photoluminescence (UCPL) and prolonged rise time for the exciton band bleach in TA measurements. By employing density functional theory (DFT) calculations, we revel the band edge energy fluctuations (~49 meV) attributed to lattice deformation in soft lattice $CsPbBr_3$, which provides additional evidence for understanding the mechanism of sub-bandgap photons absorption. We further demonstrate that the softer perovskite with a smaller Young's modulus exhibits a larger band energy fluctuation (~125 meV), signifying that the polaronic state involved optical nonlinearity can be tailored through modification of soft lattice structure in perovskite materials. This work offers valuable insights into the role of polarons in nonlinear photonics, which has significant implications for advancing optoelectronics and quantum technologies.

**Results and discussion**

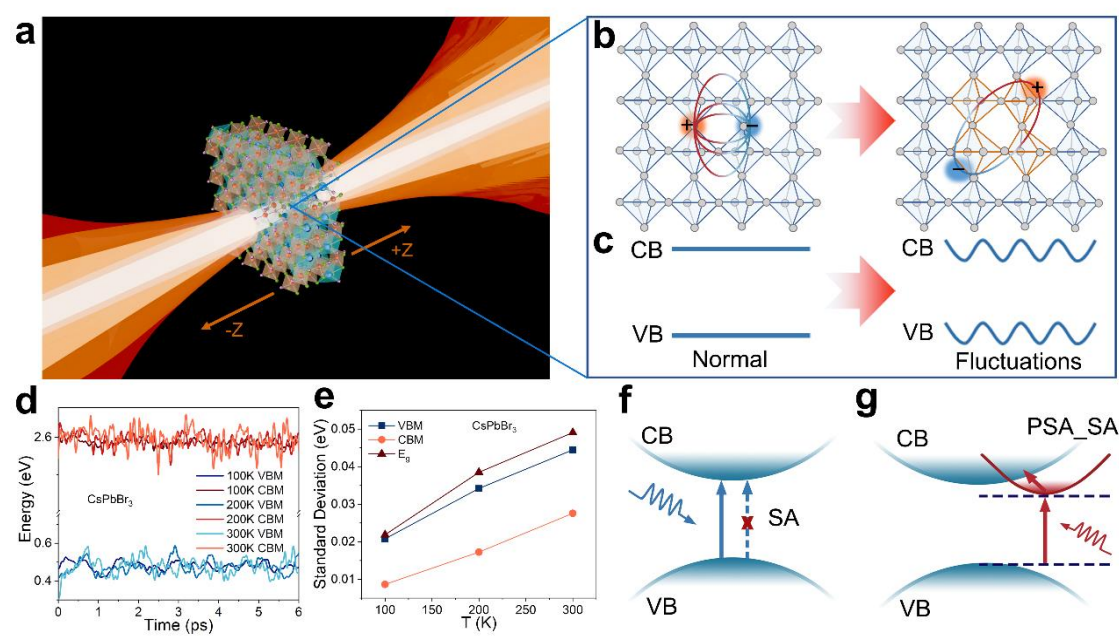

**Fig. 1| Polaronic state assisted nonlinear saturable absorption (PSA_SA) picture in LHPs. a,** Schematic diagram of Z-scan. **b,** Scheme of transformation from free excitons with columbic interaction to exciton-polaron with lattice distortion by dynamic polaronic effect. **c,** Schematic diagram of the effect of polaron formation on the band structure of materials, which will cause band energy fluctuations. **d,e,** Molecular

dynamics simulations of band energy fluctuations. The valence band maximum (VBM) and conduction band minimum (CBM) at different temperature are extracted according to the calculation results (d). The standard variations of the band energy for CBM and VBM at different temperatures (e). **f,** Schematic of SA caused by single photon absorption under up-bandgap excitation in conventional semiconductors. **g,** Schematic of PSA_SA under sub-bandgap excitation. CB and VB denote conduction band and valence band respectively.

The proposed concept of PSA_SA is illustrated in Fig. 1. In perovskite materials, when the photoexcited carriers/excitons are strongly coupled to the lattice distortion, polarizing the surrounding crystal lattice, resulting in the formation of polarons[20, 30] (Fig. 1a,b). Such polarons have a significant effect on the energy band of a functional material[20, 31], resulting in significant band energy fluctuations (Fig. 1c). To understand the complex interaction between excited electrons and the soft lattice of perovskite during polaron formation, DFT calculations were performed (Supplementary Note 2). The representative time evolution of the band edge states during 6 ps at different temperature conditions of perovskites is obtained from ab initio molecular dynamics (AIMD) simulations (Supplementary Figs.1 and 2). It can be observed that the valence band maximum (VBM) and conduction band minimum (CBM) show strong fluctuations and increase with rising temperature (Fig. 1d). The bandgap variation of $CsPbBr_3$ and $Cs_4PbBr_6$ rise from 22 meV, 57 meV at 100 K to 49 meV and 125 meV at 300 K, respectively (Fig. 1e and Supplementary Fig. 2). Similar to polaronic states driven anti-Stokes shift[29], the polarons involving energy fluctuation can break fundamental constraints of single photon transition SA (Fig. 1f and Supplementary Note 1) and enable considerable broadband SA under sub-bandgap excitation in visible region. Normally, when the photon energy is below the band gap, SA cannot be produced in principle. However, in LHPs, exciton-phonon coupling can cause significant band energy fluctuations for polariton states, which could provide enough energy for the electrons under sub-bandgap excitation to reach the conduction band, thus producing SA in Fig. 1g. Hence, we demonstrate another possible process for the existence of SA under sub-bandgap excitation, which is referred to as PSA_SA.

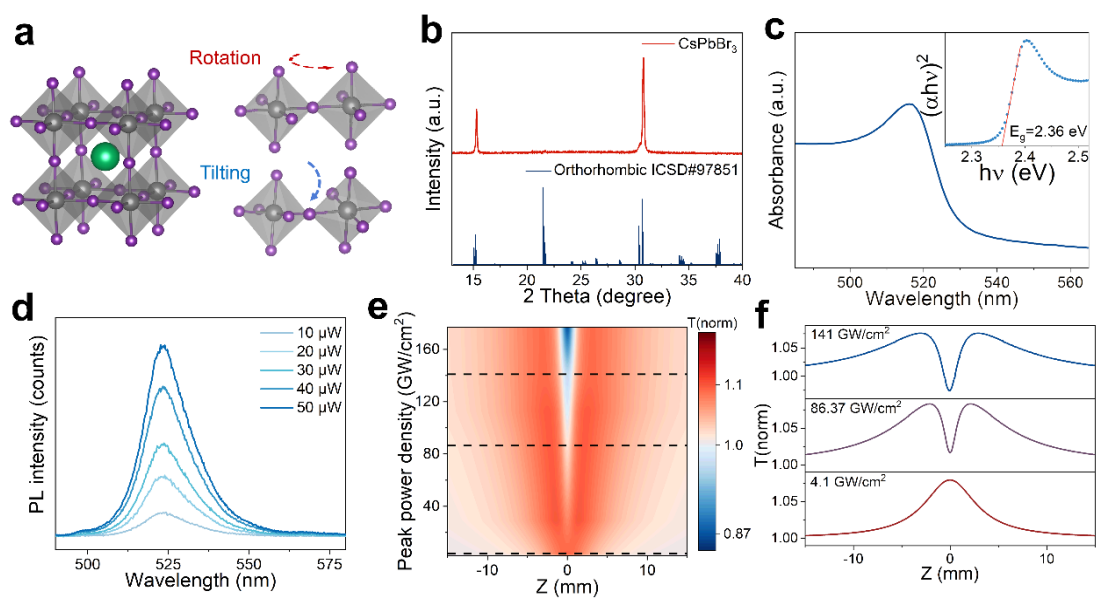

**Fig. 2| Structural, linear and nonlinear optical characterizations of CsPbBr$_3$. a,** Crystal structure of CsPbBr$_3$ and associated lattice distortion. **b,** XRD pattern of the CsPbBr$_3$ film and the standard card (ICSD-97851). **c,** UV–vis absorption spectrum and Tauc plot of CsPbBr$_3$. **d,** Steady PL spectra of CsPbBr$_3$ film. **e,** The two-dimensional map of transmittance as a function of the scan distance and the excitation peak density under 530 nm laser pulse excitation and **f,** the transmittance curves with 3 representative excitation intensities.

To achieve unconventional PSA_SA, a soft-lattice perovskite $CsPbBr_3$ was constructed. Figure 2a shows the schematic crystal structure of orthorhombic $CsPbBr_3$ and possible lattice structure distortion schemes, including rotation and tilting of the inorganic octahedral structure[32-34]. The $CsPbBr_3$ film was prepared on the sapphire substrate in a chemical vapor deposition (CVD) system (See details in Supplementary Note 3). The scanning electron microscopy (SEM) image shows that the surface of the film is flat and smooth without any grain boundaries or cracks and energy dispersive X-ray spectroscopy (EDS) elemental mappings reveal the uniform spatial distribution of Cs, Pb and Br elements (Supplementary Fig. 3). The atom force microscopy (AFM) shows that the thickness of $CsPbBr_3$ film is 184 nm and the surface roughness is less than 5 nm (Supplementary Fig. 4). The X-ray diffraction (XRD) patterns in Fig. 2b with peaks at ~ 15.1°, 15.3°, 30.5° and 30.8° correspond to (002), (110), (004) and (220) planes of orthorhombic $CsPbBr_3$ (ICSD 97851), respectively. Meanwhile, no other peaks are observed, further indicating the high crystalline purity of $CsPbBr_3$ film. As shown in Fig. 2c, the film exhibits an absorption peak at ~516 nm, implying obvious excitonic absorption. The Tauc plot calculated from the absorption spectra shows a bandgap of 2.36 eV (Fig. 2c and Supplementary Note 4). In addition, the PL spectra show a single symmetric emission center at 524 nm (2.37 eV) with the full width of maximum (FWHM) ~16 nm (Fig. 2d), which is consistent with previous report[35]. We investigated the structural stability of $CsPbBr_3$ film, the XRD signals of $CsPbBr_3$ film stored in air for 4 months are almost consistent with the initial thin film (Supplementary Fig. 5).

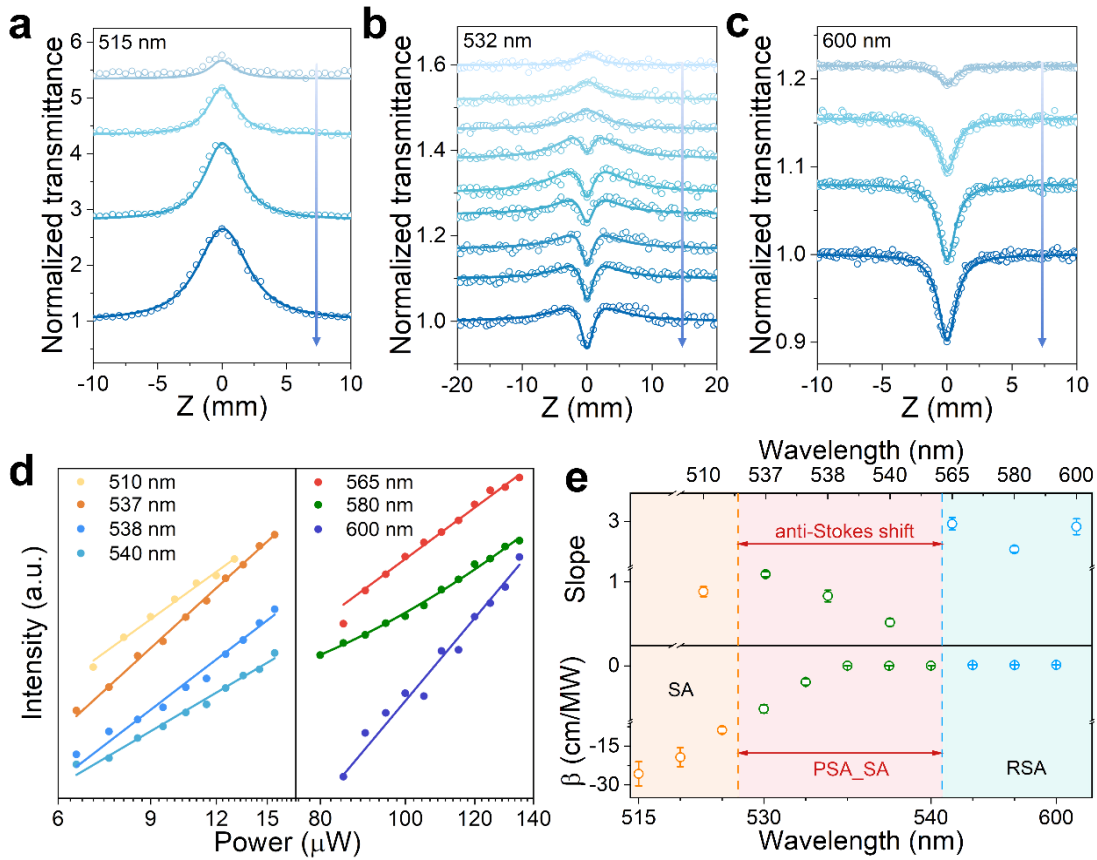

**Fig. 3| Z-scan and steady-state PL studies of polaronic state assisted nonlinear saturable absorption (PSA_SA).** The excitation wavelengths are **a,** 515 nm (pump intensity: 2.07-4.25 GW/cm$^2$), **b,** 532 nm (pump intensity: 3.24-94.86 GW/cm$^2$), and **c,** 600 nm (pump intensity: 48.60-215.14 GW/cm$^2$), respectively. From top to bottom, the light intensity increases in turn. **d,** Power dependence of the PL intensity at different excitation wavelengths for CsPbBr$_3$. **e,** Wavelength dependence of slope (top) and nonlinear absorption coefficient β (bottom) for CsPbBr$_3$. Error bars represent the standard error.

The nonlinear absorption properties of CsPbBr$_3$ film are characterized by femtosecond Z-scan technique (see Supplementary Note 5 for details). The normalized excitation intensity-dependent optical transmittance under sub-bandgap excitation (2.34 eV) is presented in Fig. 2e, which reveals that the transmittance of CsPbBr$_3$ film is distinct with increasing excitation intensity. At low light intensity, the highest transmittance appears at the focal point, which indicates SA, and when the light intensity gradually increases, the typical reverse saturation absorption (RSA) feature gradually appears with a transmittance valley near the focal point. We observe distinct SA in this perovskite under sub-bandgap excitation, as well as the characteristic nonlinear absorption conversion from SA to RSA with increasing excitation power from low to high intensity (Fig. 2f). Such anomalous SA does not originate from single-photon transition optical bleaching, owing to the insufficient photon energy provided by sub-bandgap excitation.

To elucidate the underlying mechanism of aforementioned anomalous SA phenomenon, the nonlinear absorption responses of $CsPbBr_3$ film were collected within the wavelength range of 515 nm-600 nm (Supplementary Fig. 6). When the excitation photon energy exceeds optical bandgap (515-525 nm), the $CsPbBr_3$ film behaves as SA (Fig. 3a and Supplementary Fig. 6a-c). This behavior is attributed to the absorption change caused by electrons transition to the conduction band through single-photon absorption (schematics in Fig. 1f). When the excitation photon energy is slightly below the bandgap (excitation wavelength within 530-540 nm), both SA and RSA are observed in the $CsPbBr_3$ film (Supplementary Fig. 6d-h), with a conversion from the SA to RSA as the laser intensity increases (Fig. 3b). It is reasonable to deduce that such a nonlinear absorption arises from the competition between two nonlinear absorption mechanism. As the excitation photon energy decreases further below the bandgap, that is, the corresponding excitation wavelength is longer than 540 nm, the SA ceases to occur, and complete RSA becomes distinctly observable (Fig 3c and Supplementary Fig. 6i-k). In principle, when the excitation photon energy is below the band gap, SA theoretically does not occur because the photon energy is insufficient to excite electrons from the valence band into the conduction band. Although two-photon absorption is energetically feasible in this case, the probability of this process occurring at low light intensity is extremely low. Combined with the soft lattice and strong electron-phonon coupling properties of LHPs[36, 37], exciton-polaron induced energy fluctuation, i.e. polaronic state, may involve in this unconventional SA process. In other words, electron-phonon coupling induced exciton-polaron facilitate the transition of low-energy electrons to the conduction band, thereby leading to the occurrence of the SA phenomenon (Supplementary Fig. 6d-e). As the excitation photon energy gradually moves away from the band gap, the polaron energy is no longer sufficient to support the complete SA. At this time, the contribution of two-photon absorption becomes the dominant, but the SA phenomenon can still be observed, as shown in the Supplementary Fig. 6f-h. When the excitation photon energy (550-600 nm) is much smaller than the optical band gap, the $CsPbBr_3$ film exhibits a complete RSA (Fig. 3c and Supplementary Fig. 6i-k). To determine whether the RSA at these wavelengths is caused by the two-photon absorption, the relationship between Ln (1-T) and Ln (I) is fitted. The fitting results show that k=1.1 (Supplementary Fig. 7), which indicates the occurrence of a two-photon absorption dominated nonlinear RSA process[38].

To verify the PSA_SA behavior under sub-bandgap excitation, we conducted PL and ultrafast TA spectra measurements of $CsPbBr_3$ film. The PL spectra of $CsPbBr_3$ excited by photon up-conversion and down conversion processes at room temperature are shown in Supplementary Fig. 8. In order to reveal the basic mechanism of PL process, the excitation power dependence of the PL intensity at each excitation wavelength is investigated. The power dependence of the PL intensity in a log-log scale for all excitation wavelengths was shown in Fig. 3d. The measured power-dependent PL intensities are fitted with the power law[39] $I = \alpha P^S$, where $P$ is the excitation power, $\alpha$ is the fitting parameter, and $S$ is the exponent of the power law which represents the slop of the curve. The extracted values of $S$ as a function of the corresponding excitation wavelength are further plotted in Fig. 3e (up panel). When the excitation wavelength $\lambda_{exc}$ ≤540 nm, both down conversion and up-conversion PL present an approximate linear or sublinear trend with the excitation density, indicating that the PL originates from exciton radiative recombination. Combined with the strong electron-phonon coupling properties, the observed linear and sublinear relationships between UCPL intensity and excitation power of $CsPbBr_3$ film suggest that UCPL follows a one-photon up-conversion process involving multi-phonon-assisted[40]. The anti-Stokes shift is defined as the energy difference between the excitation and emission peak photon energy. Thus, the anti-Stokes shift corresponding to phonon-assisted UCPL in $CsPbBr_3$ is about 70 meV (Supplementary Fig. 9a). When the excitation wavelength $\lambda_{exc}$≥565 nm, the slope of UCPL is almost three times correlated with the excitation density dependence. This may be due to nonlinear optical effects caused by multiple-photon absorption or Auger recombination[41, 42], because these mechanisms typically display a superlinear (quadratic or more) excitation power dependence.

In order to clarify the SA phenomenon below the bandgap excitation more directly, we calculate the nonlinear absorption coefficients of $CsPbBr_3$ at different excitation wavelengths, as shown in Fig. 3e (down panel). According to the nonlinear absorption characteristics and the nonlinear absorption coefficient, the excitation wavelengths are divided into three regions: the SA region caused by pure single photon absorption when the excitation photon energy exceeds the band gap ($\lambda_{exc} \leq 525$ nm); the RSA region which caused by two-photon absorption when the excitation photon energy is significantly lower than the bandgap ($\lambda_{exc} \geq 550$ nm); and the region in which SA and RSA coexist when the excitation photon energy marginally less than the bandgap (530 nm $\leq \lambda_{exc} \leq$ 540 nm). The energy shift is defined as the difference between the excitation photon energy corresponding to the SA effect still existing in $CsPbBr_3$ under sub-band gap excitation and the optical band gap, which is about 60 meV (Supplementary Fig. 9b). The nonlinear optical response of $CsPbBr_3$ film depending on excitation photon energy shows high consistency with corresponding PL results. The single photon fluorescence region corresponds to the SA region; the region where SA and RSA coexist is consistent with the region of one-photon up-conversion involving phonon-assisted; and the multiphoton fluorescence region corresponds to the RSA region. Notably, the phonon-involved anti-Stokes shift invariably overlaps with the sub-bandgap excitation SA occurrence band region, providing strong evidence for excitation-polarons-assisted SA. In addition, we also observe a similar phenomenon in another perovskite ($Cs_4PbBr_6$). The PL spectra of $Cs_4PbBr_6$ at different excitation wavelengths are shown in Supplementary Fig. 10. The power dependence of the PL intensity for all excitation wavelengths and the extracted values of *S* as a function of the corresponding excitation wavelength are shown in Supplementary Fig. 11a, b (up panel). Since the emission center of $Cs_4PbBr_6$ is at 517 nm, the phono-assisted anti-Stokes shift obtained from the experimental results is about 102 meV. According to the absorption spectrum, it can be observed that $Cs_4PbBr_6$ has an absorption edge at 517 nm (Supplementary Fig. 12a) and the optical band gap calculated by Tauc method is about 2.32 eV (Supplementary Fig. 12b). Combined with the Z-scan results (Supplementary Fig. 11b (down panel)), it is obtained that the energy shift is about 70 meV. The energy shift of $Cs_4PbBr_6$ is larger than that of $CsPbBr_3$, which is due to the fact that $Cs_4PbBr_6$ with decoupled octahedra is susceptible to perturbation by photoexcitation, while $CsPbBr_3$ has a rather rigid inorganic framework[43]. In order to prove this speculation, we calculate the Young's modulus of two perovskites by DFT (see details in Supplementary Note 6). Different models are used to calculate the Young's modulus for each system (Supplementary Table 1). The results show that the Young's modulus of $Cs_4PbBr_6$ (16.52 Gpa) is smaller than that of orthorhombic $CsPbBr_3$ (17.98 Gpa). Apart from the mean value of Young's modulus, the calculated anisotropy of Young's modulus (Supplementary Table 2 and Supplementary Fig. 13) further demonstrate that $Cs_4PbBr_6$ exhibits a "softer" lattice feature compared with $CsPbBr_3$. The bandgap variation of $Cs_4PbBr_6$ increases from 57 meV (100 K) to 125 meV (300 K), indicating that the energy band fluctuation of LHPs with softer crystal lattice is more significant. These calculation results demonstrate that $Cs_4PbBr_6$ with "softer" lattice facilitates stronger electron-phonon coupling during light-matter interaction. In addition, these results also show that the optical nonlinearity can be regulated by adjusting the lattice structure.

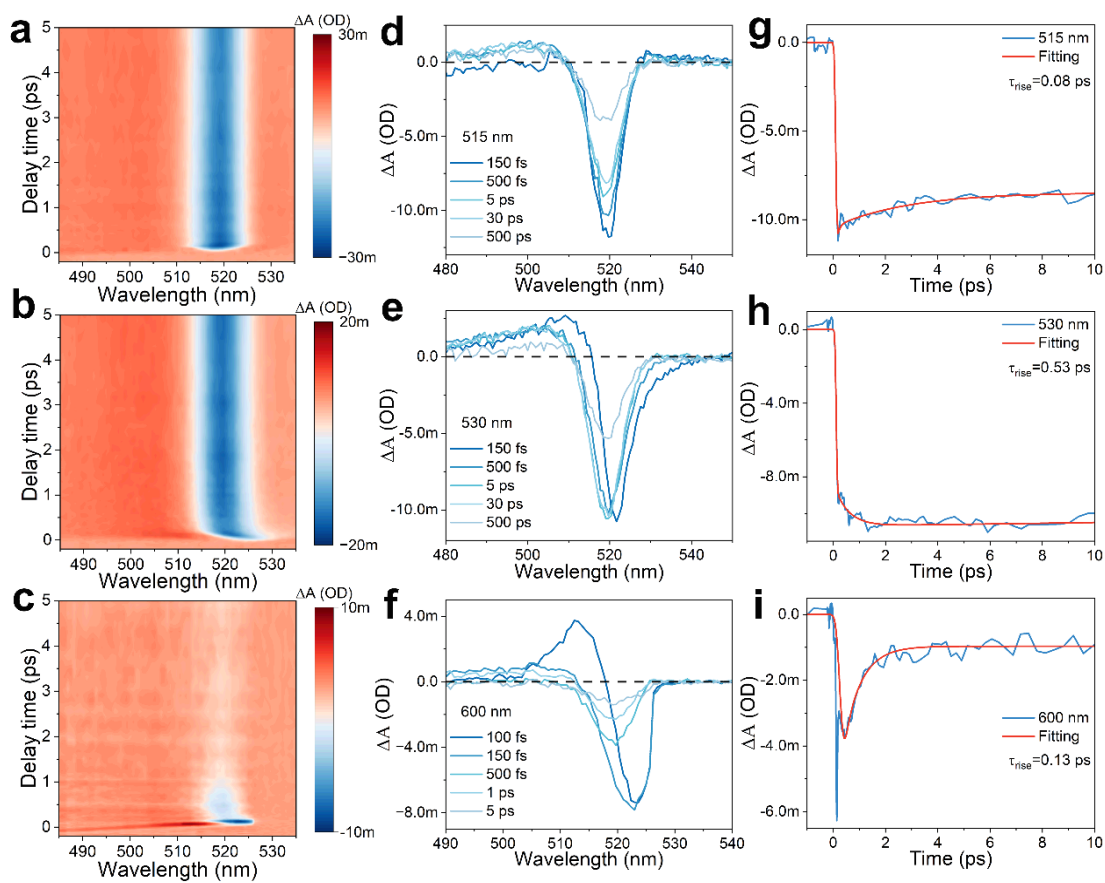

**Fig. 4| Transient absorption (TA) studies confirming polaronic state assisted nonlinear saturable absorption (PSA_SA). a-c,** Representative pseudo-color TA spectrums at 515 nm (**a**), 530 nm (**b**) and 600 nm (**c**) excitations, respectively. **d-f,** Time-dependent photo-induced changes in absorption (ΔA) which is pumped by 515 nm (**d**), 530 nm (**e**) and 600 nm (**f**), respectively. **g-i,** Exciton band bleach dynamics monitored at 519 nm for 515 nm (**g**), 530 nm (**h**), and 600 nm (**i**) excitations, respectively.

In order to further reveal the mechanism of PSA_SA in CsPbBr$_3$ film under sub-bandgap excitation, we perform systematic ultrafast TA measurements (Supplementary Note 7) of CsPbBr$_3$. Figure. 4a-c show the pseudo-color plot of TA difference (ΔA) as a function of probe light wavelength and delay time at different excitation wavelengths.

As shown in Fig. 4d, when photoexcitation energy is larger than the band gap (515 nm), the TA profile displays a negative peak at the exciton resonance (ΔA<0) and two positive peaks at either side (ΔA>0), in which the negative signal represents photobleaching and the positive signal represents photo-induced absorption. These features are a synthesis of the state filling of the exciton band and the effects of exciton band broadening and shifting caused by multibody interactions in the presence of excitons and carriers[29, 44]. When the excitation photon energy (530 nm) is slightly less than the band gap, the TA spectra show a slight blueshift (Fig. 4b, e). After this slight blueshift, the transient spectral characteristic for the excitation below the gap is similar to that for the excitation above the gap. When the photon energy (600 nm) is much smaller than the optical band gap, the TA spectra show an instantaneous giant response at time zero corresponding to the blueshift of excitation (Fig. 4c, f). This is due to the optical Stark effect caused by the hybridization of the equilibrium and Floquet states[29, 45]. In addition, the transient dynamics monitored at the exciton band (~519 nm) under different excitation wavelengths are extracted (Fig. 4g-i and Supplementary Fig. 14). The rise time for the exciton band bleaching under 515 nm excitation is around 0.08 ps (Fig. 4g), which corresponds to the time required for the excitations to scatter to the band edge by optical phonons[29, 46]. For 530 nm excitation, the rising of exciton band bleach is relatively slower with a time constant of 0.53 ps (Fig. 4h). When the excitation photon energy is much less than the band gap, two-photon absorption will dominate due to the larger pump energy. After removing the strong nonlinear response by optical Stark effect, the rise time of exciton band bleach is around 0.13 ps (Fig. 4i). The corresponding relationship between the rise time of exciton band bleaching and excitation wavelength is shown in Supplementary Fig. 15. When the excitation photon energy is larger than the optical band gap (515-525 nm), the average rise time of this band is almost 0.08 ps; when the photon energy is slightly less than the band gap (530-540 nm), the average rise time is around 0.60 ps; when the photon energy is much less than the band gap (550-600 nm), the average rise time is around 0.15 ps. When the excitation photon energy is slightly less than the band gap, the rise time of exciton band bleaching is slower which indicated strong electron-phonon interaction occurs, causing the emergence of excitation polarons[29, 31]. The results of TA also consistent with PL and Z-scan results, further confirming that the SA under sub-bandgap excitation is assisted by exciton-polarons induced polarons state.

In conclusion, we have investigated the nonlinear absorption properties of LHPs at room temperature and revealed the PSA_SA behavior under sub-bandgap excitation. The DFT calculations show that the strong fluctuations of band energy caused by the rapid lattice deformation and formation of polarons, which leads to the substantial energy shift. The polarons involving energy fluctuation assistance can break the basic restriction of single-photon transition and provide sufficient energy for single-photon absorption under sub-bandgap excitation. We have further identified that the perovskite with "softer" lattice exhibits greater band energy fluctuation and facilitates stronger carrier/excitons-phonon coupling induced polarons. Our findings reveal that modulating the lattice properties of LHPs may be a way of adjusting the polaron states to broaden nonlinear absorption applications.

**Data availability**

Data that support the plots within this paper and other findings of this study are available from the corresponding authors upon reasonable request. Source data are provided with this paper.

**Acknowledgements**

This research was supported by the National Natural Science Foundation of China (Nos. 62275275, 11904239, 62422506, 12474383, 52273202), National Key R&D Program of China (2022YFA1604200), Natural Science Foundation of Hunan Province (Grant Nos. 2021JJ40709, 2022JJ20080, 2024JJ6481) and Postgraduate Innovative Project of Central South University (Grant No. CX20230246). This work was also supported in part by the High-Performance Computing Center of Central South University and Open Sharing Found for the Large-scale Instruments and Equipment of Central South University. We also acknowledge resources from the Hefei Advanced Computing Center.

**Author contributions**

Y. W. Wang and J. He conducted and supervised the project. L. Zhou fabricated the perovskite film, conducted a series of data collection and analysis. Y. D. Wang and C. Tong carried out the density functional theory calculations. J. Kang, X. Li, Q. Long and X. Zhong helped with the optical tests. Z. Chen carried out the photoluminescence

**Competing interests**

The authors declare no competing interests.